\documentclass[a4paper]{jpconf}
\usepackage{iopams}
\usepackage{graphicx}
\begin{document}

\title{Slicing the Torus: Obscuring Structures in Quasars}
\author{Martin Elvis$^1$}
\address{$^1$ Harvard-Smithsonian Center for Astrophysics, 60 Garden
  St. Cambridge MA02138 USA } 

\ead{elvis@cfa.harvard.edu}

\begin{abstract}
Quasars and Active Galactic Nuclei (AGNs) are often obscured by dust and gas. It
is normally assumed that the obscuration occurs in an oblate "obscuring torus",
that begins at the radius at which the most refractive dust can remain
solid. The most famous form of this torus is a donut-shaped region of molecular
gas with a large scale-height. While this model is elegant and accounts for many
phenomena at once, it does not hold up to detailed tests. Instead the
obscuration in AGNs must occur on a wide range of scales and be due to a minimum
of three physically distinct absorbers. Slicing the "torus" into these three
regions will allow interesting physics of the AGN to be extracted.

\end{abstract}

%%%%%%%%%%%%%%%%%%%%%%%%%%%%%%%%%%%%%%%%%%%%%%%%%%%%%%%%%%%%%%%%%%%%%%%%
\section{The Quasar Standard Model}

There is a "standard model" for quasars, which was put in place within a decade
of the discovery of quasars \cite{1963Natur.197.1040S}, i.e. by 1973. This
standard model consists of three elements: (1) a {\em supermassive}
(10$^6$-10$^9$ M$_{\odot}$) {\em black hole} (SMBH, \cite{1969Natur.223..690L}),
surrounded by (2) an {\em accretion disk} \cite{1973A&A....24..337S}, with (3) a
{\em relativistic jet} \cite{1966Natur.211..468R, 1974MNRAS.169..395B} emerging
perpendicular to the disk and originating at just a few Schwarzchild radii away
from the black hole. The elements of this model successfully account for, in
turn: (1) the total power output of the quasar, from the gravitational energy
released by infall to near the event horizon of the black hole; (2) the maximum
temperature of 50,000~K - 100,000~K of the ultraviolet (UV) continuum that
dominates the luminosity, from the thermalization of the gravitational energy
release in the accretion disk; and (3) the phenomenology of apparent
superluminal motion, rapid variability and polarization of the radio emission in
those quasars where a jet is pointing almost directly at us (the blazars). This
is a pretty good list of successes, and they have held up well against decades
of tests.

However, unlike the predictive power of the contemporaneous particle physics
standard model \cite {1967PhRvL..19.1264W}, the quasar standard model predicts
little of the rich phenomenology of quasars: (a) the various 'types' of active
galaxies \cite{1974ApJ...192..581K, 1981ApJ...249..462O}; (b) the maximally hot
dust found in quasars, but not in starburst galaxies \cite{1978ApJ...226..550R,
1994ApJS...95....1Eshort, 2006ApJ...640..579G}, (c) the strong X-ray emission
\cite{1978MNRAS.183..129E, 1979ApJ...234L...9Tshort}, and (d) all of the many
atomic emission and absorption features seen in the spectra of quasars
\cite{1997iagn.book.....P}, not even the broad ($\sim$1\% $c$) emission lines
that led to their initial recognition as exceptional objects
\cite{1943ApJ....97...28S}.

As a response to several of these gaps, a fourth element of the standard model
has been commonly accepted since about 1985: a flattened, but large
scale-height, {\em obscuring torus}.

%%%%%%%%%%%%%%%%%%%%%%%%%%%%%%%%%%%%%%%%%%%%%%%%%%%%%%%%%%%%%%%%%%%%%%%%
\section{The Obscuring Donut Torus} 

A flattened obscuring torus coaligned with the accretion disk is an appealingly
simple addition to the standard model. It is able to explain both the variety of
types of AGNs, the existence of maximally hot dust, and the bi-conical
morphology often shown by the narrow emission lines of AGNs. This paradigm is
known as the Unified Scheme \cite{1982ApJ...256..410L, 1989ApJ...336..606B,
1988ApJ...329..702K, 1993ARA&A..31..473A, 1995PASP..107..803U}. The Unified
Scheme posits obscuration by an optically and geometrically thick "torus", lying
between the inner, broad ($\sim$10$^4$ km~s$^{-1}$), and outer ($\sim$10$^3$
km~s$^{-1}$), narrow, line emitting regions.
% Certainly, some non-spherical obscuring region clearly creates several of
% these effects in AGNs. 

The torus is almost universally taken to be a large scale-height (H/R$\sim$1),
cold structure, rich in molecular gas and dust, that is co-planar with the
accretion disk \cite{1988ApJ...329..702K}. As such it resembles a donut with a
hole in the center, notably in the famous illustration in Urry \& Padovani
\cite{1995PASP..107..803U}. This specific form of non-spherical obscuring region
has been called the "donut torus" \cite{2006ApJ...648L.101E}. The "strong" form
of the Unified Scheme asserts that our orientation relative to the one
jet/disk/torus axis explains {\em all} of the variety of AGN types.

The strong Unified Scheme cannot be 100\% correct as, e.g.  the incidence of
X-ray obscuration is clearly more common at low luminosities
\cite{1998ApJS..117...25M, 2008ApJ...675..960P, 2010ApJ...714..561L,
2011ApJ...728...58B}, requiring some modification \cite{1991MNRAS.252..586L}.

Nontheless this scheme does explain \cite{1982ApJ...256..410L,
1985ApJ...297..621A}: (1) the distinction between type~1 (with broad emission
lines) and type~2 (no broad emission lines) AGNs, due to optical dust
obscuration between the two emission regions; (2) the presence of heavy X-ray
obscuration in type~2 AGNs, due mainly to gas; (3) the biconical geometry of the
outer narrow line region, due to geometric collimation of the continuum; (4) the
finding of polarized broad lines in otherwise purely narrow-lined type~2 AGNs
\cite{1985ApJ...297..621A, 1990ApJ...355..456M, 1995ApJ...440..597T}, due to
scattering off warm electrons above the torus; and (5) the relative space
densities of type~1 and type~2 AGNs. All these achievements apply equally to
radio-loud AGNs \cite{1989ApJ...336..606B}, and can be successfully extended to
connect the luminosity functions of blazars and radio galaxies
\cite{1991ApJ...382..501U}.

This is a long list of accomplishments, and there can be no doubt that flattened
obscuring regions are important in AGNs. However, the elegant reduction of these
effects to a single region cannot be sustained for both theoretical and
observational reasons.

%%%%%%%%%%%%%%%%%%%%%%%%%%%%%%%%%%%%%%%%%%%%%%%%%%%%%%
\subsection{Problems with the Donut Torus}

There are four theoretical challenges to the donut torus picture of AGN
obscuration. While none of them is individually inescapable, together they are a
significant challenge. They are:
\begin{enumerate}
\item {\em The large, H/R$\sim$1, scale height in cold material:} Clearly a cold
($\sim$100~K) medium does not have thermal velocities of greater than the
$\sim$1000~km~s$^{-1}$ of the narrow emission lines, as would be required to
reach H/R$\sim$1 interior to the region emitting those lines. (This assumes that
the lines have widths comparable to virial or Keplerian values.) The alternative
is that the material is highly clumped and has many clouds on highly inclined
orbits. A clumpy torus explains the observed AGN spectral energy distributions
(SEDs) better than a continuous medium \cite{2008ApJ...685..160N}.  A clumpy
torus is observationally supported by the emission seen outside of the bicones
in NGC~4151, both in optical emission lines (the "rogue clouds" of
\cite{2005AJ....130..945Dshort}) and soft X-rays
\cite{2011ApJ...742...23Wshort}. However these clouds must then collide with one
another and collapse on some unclear, but probably short, timescale. A
non-static, though possibly steady-state, model is needed
(e.g.\cite{2004A&A...413..949V, 2005ApJ...630..167T}).

\item {\em The energy and photon deficit problem for the broad emission line
region}: Netzer (1987) noted that the observed UV continuum could not be
extrapolated in a simple, accretion disk like, way
(e.g. \cite{1987ApJ...323..456M}) and still provide sufficient photons to ionize
the gas producing the broad emission lines. More recent inferred continuum
shapes suggest even fewer photons in the extreme UV \cite{1997ApJ...477...93L,
1997ApJ...475..469Z}. Nor does the continuum provide sufficient energy input to
power the broad emission lines \cite{1993PASP..105.1150B}. Netzer's solution was
to locate the broad emission line region directly above the accretion disk. In
this way the broad emission line gas sees a more powerful continuum than an
observer at a random angle. This explanation works as, above the disk, neither
the geometrical cosine~$\theta$ nor limb darkening \cite{1985A&A...143..374S}
diminish the continuum. However, if the torus is co-planar with the disk, and
covers the direct view of the disk over 80\% of side-on viewing directions, then
we can only see the broad emission lines almost perpendicular to the disk, so we
also see the same continuum as that gas. The continuum may not be a simple
extrapolation of an $\alpha$-disk spectrum, but could have a second,
short-wavelength, peak, e.g. due to blurred Lyman-$\beta$ and HeII emitted by
dense clouds at small radii Lawrence \cite{2011arXiv1110.0854L}, which may solve
these problems.

\item {\em The impedence of feedback:} As is now well known, the mass of the
stellar bulge and the mass of the SMBH in a galaxy are closely tied together
(see, e.g. \cite{2005SSRv..116..523F}). Assuming that the two masses are
causally linked (but see \cite{2011ApJ...734...92J}), some feedback
mechanism must keep the two growing in synchrony. There are three ways the power
output of a quasar can interact with the host galaxy interstellar medium (ISM):
(1) radiation; (2) broad slow winds; (3) narrow relativistic jets. While
relativistic jets from their central galaxies are clearly important in clusters
of galaxies \cite{2007ARA&A..45..117M}, this mechanism is unlikely to be
important in either less massive systems with lower density ISMs, or in the 10
times more numerous radio-quiet quasars. However, both radiation and slow winds
will be inhibited by the presence of an exterior donut torus that blocks 80\% of
the sky as seen from the inner nucleus. This large covering factor increases the
efficiency needed to deposit the $\sim$5\% of the total radiative quasar power
needed to unbind the host galaxy ISM \cite{2005Natur.433..604D,
2006ApJS..163....1H} to $\sim$25\% of the {\em escaping} radiation. More
efficient scenarios have been proposed however \cite{2010MNRAS.401....7H}.

\item {\em The wide range of obscuring X-ray column density:} The donut torus
was invoked to explain AGNs such as NGC~1068, in which the direct central
radiation was completely blocked by a Compton thick obscurer [$\tau_{Compton}$
=1 corresponds to N$_H$=2$\times$10$^{24}$cm$^{-2}$]. However, there are many
intermediate type AGNs, with a wide range of lower X-ray column densities, and
optical reddening values. Values of N$_H$ from $\sim$10$^{21}$cm$^{-2}$ to
$>$10$^{25}$cm$^{-2}$ are often observed. This is roughly the difference between
a thin sheet of paper and a brick, so that a single physical cause is not
required, and may be hard to produce. Usually, in the donut torus scheme, these
intermediate column densities are attributed to viewing the donut torus at a
grazing angle \cite{1988ApJ...327...89O}. In the case of NGC~4151 this may be
correct \cite{2005AJ....130..945Dshort}. However, these intermediate column
densities are rather common \cite{1999ApJ...522..157R}.
%
%E.g. in the "Piccinotti sample" (selected by 2-10 keV flux,
%\cite{2005A&A...432...15P}, there are XX type 1.8-1.9 AGNs compared to XX type
%1-1.5 AGNs. 
%
The extended "atmosphere" of the donut torus must then cover $\sim$35\% of the
sky seen from the inner nucleus, which is a substantial change to the picture. A
clumpy wind can produce the observed N$_H$ distribution
\cite{2008ApJ...685..160N}.
\end{enumerate}

One recurring {\em observational} problem is that the broad line widths seem to
be dominated by orbital rotation \cite{1986ApJ...302...56W, 2005MNRAS.359..846S,
2007Natur.450...74Y}. That requires that most broad emission line regions are
seen at a large angle to the rotation axis, but the donut torus would prevent
this.

These problems can be eased if one or more of the assumptions of the strong
Unified Scheme are relaxed: if the ratio of type~2:type~1 AGNs is smaller than
had been thought, if the torus is not co-planar, if the X-ray and optical (or
equivalently the gas and dust) obscurers are not tightly connected, or if the
obscuring region can be sliced into several layers. I will discuss this last
possibility next.

%%%%%%%%%%%%%%%%%%%%%%%%%%%%%%%%%%%%%%%%%%%%%%%%%%%%%%%%%%%%%%%%%%%%%%%%
\section{Slicing the Torus} 

In the years since the donut torus was proposed there has been much detailed
observational work on each of the features that the scheme was invented to
explain. One consequence is accumulating evidence that obscuration in AGNs
occurs on both smaller and larger scales than that of the hot dust limited
radius of the donut torus alone. Here I slice the obscuration into three scales
(Figure~1).

%%%%%%%%%%%%%%%%%%
\begin{figure}
\begin{center}
\includegraphics[width=11cm]{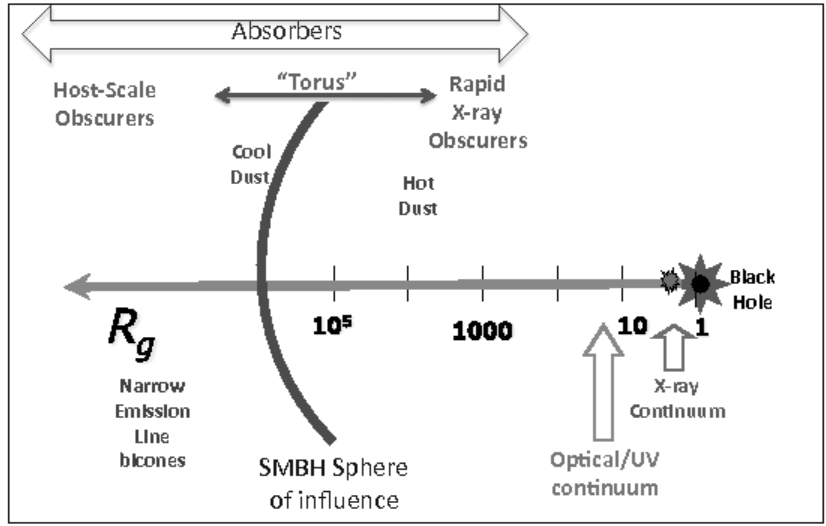}
\end{center}
\caption{\small Scales of obscuration in quasars and AGNs. The horizontal scale
  is in log(R$_g$).}
\end{figure}
%%%%%%%%%%%%%%%%%%

%%%%%%%%%%%%%%%%%%%%%%%%%%%%%%%%%%%%%%%%%%%%%%%%%%%%%%
\subsection{Smaller Scale Obscuration}

Large amplitude changes in absorbing X-ray column density have been seen on
briefer and briefer time intervals. There are now numerous examples of large
changes ($\Delta$N$_H>$10$^{23-24}$cm$^{-2}$) in a few days or hours
(e.g. \cite{2004ApJ...615L..25E, 2007MNRAS.377..607P}). The most dramatic
example is NGC~1365, in which the X-ray source underwent a total eclipse by a
Compton-thick cloud within two days, and emerged within another two days
\cite{2007ApJ...659L.111R}. For any velocity comparable to the Keplerian
velocity for the eclipsing clouds, these events all imply that they lie at about
the same distance as the broad emission line gas ($r<$10$^{3-4}~R_S$) and have
comparably high densities ($n_e>$10$^{9-10}$cm$^{-3}$). The obvious conclusion
is that the eclipsers are discrete clouds of broad emission line gas. Further
investigations seem to be revealing unexpected features in these clouds with
surprising implications \cite{2010A&A...517A..47M, 2012arXiv1201.3520E}.

For our purpose here, these eclipses show that obscuration, at least by gas,
occurs well within the dust sublimation radius. The observed low dust-to-gas
ratios, from comparisons of optical reddening with X-ray absorption
\cite{1982ApJ...257...47M, 2001A&A...365...28M}, may be explained by a mix
of inner, dust-free, and outer, dusty, absorbers \cite{2002ApJ...571..234R}.

As the UV continuum source has $\sim$10 times the radius and so $\sim$100 times
the area of the X-ray continuum, it is less likely that similar large amplitude
eclipses could be seen at UV wavelengths, which makes it hard to tell if the
clouds are dusty, as suggested by Czerny \& Hryniewicz
\cite{2011A&A...525L...8C}. (In principle transiting exoplanet techniques could
look for these partial eclipses.)

%%%%%%%%%%%%%%%%%%%%%%%%%%%%%%%%%%%%%%%%%%%%%%%%%%%%%%
\subsection{Larger Scale Obscuration}

Since 1980 there have been many papers demonstrating a clear connection between
the obscuration of the AGN and the inclination angle of the host galaxy
\cite{1980AJ.....85..198K, 1982ApJ...256..410L, 1985A&A...147....1D,
1988ApJ...330..121T, 1990AJ.....99.1722K, 1990AJ.....99.1435K,
1997ApJ...489..615S, 2006ApJ...645..115R}. Evidently there is some obscuring
region dominated by the host galaxy dynamics, and so outside the SMBH sphere of
influence, though the radial scale of this obscurer is uncertain. Moreover, as
the radio jets of these galaxies lie at random angles to the host galaxy disk
plane, within a very wide cone \cite{2000ApJ...537..152K, 2003ApJS..148..327S},
the obscurer, the jet and the accretion disk cannot all share the same axis.

Hubble imaging of type~1 and type~2 hosts also clearly shows that there are many
more dust lanes crossing the nuclei of type~2 AGNs than of type~1s on a several
100~pc scale \cite{1998ApJS..117...25M}. This obscuration lies beyond the region
producing the infrared emission from these AGNs. As these type~2 hosts are of
later morphological type than the type~1s, they undermine the strong Unified
Scheme.

A different argument against the donut torus, coming from properties on
similarly large scales of up to $\sim$1~kpc , is based on the bi-conical regions
marked out by the narrow emission lines. A series of papers using Hubble STIS
long slit spectroscopy \cite{2000ApJ...532L.101C, 2011ApJ...727...71F} has found
that these bi-cones are not made up of pre-existing ISM illuminated by the AGN,
as in the donut torus model, but by outflowing hollow cones with wide opening
angles. Hence the bi-cone shapes are often not formed by a collimated incident
radiation pattern, but are matter-bounded, i.e. limited by the shape of the wind
coming from the nucleus. Moreover, the wide angles derived for the wind bi-cones
($\sim$50$^{\circ}$ \cite{2010AJ....140..577F}) imply that any torus covers a
more limited solid angle than in the donut torus model. (These bi-cones also
show feedback in action, \cite{2011ApJ...742...23W}).

Larger scale obscuration may be due to matter streaming toward the accretion
disk, as observed in H$_2$ emission \cite{2010MNRAS.402..819S,
2011MNRAS.417.2752R}, and predicted in some galaxy merger models
\cite{2009A&A...495..137H}.  Improved imaging in molecular lines in the mm- and
sub-mm bands, primarily of CO, have begun to image obscuring structures at the
$\sim$100~pc radius scale. In the prototypical type~2 AGN, NGC~1068 (for which
the donut torus was largely originally created \cite{1985ApJ...297..621A}), a
Compton thick warped disk with an inner radius of $\sim$70~pc curls up to just
cross our line of sight to the nucleus \cite{2000ApJ...533..850S}. This
molecular disk appears to be disconnected from any smaller structure stretching
down to the dust sublimation radius. ALMA will revolutionize this field.

%%%%%%%%%%%%%%%%%%%%%%%%%%%%%%%%%%%%%%%%%%%%%%%%%%%%%%
\subsection{Hot Dust Scale Obscuration}

Nonetheless, a substantial fraction of the AGN nuclear power is absorbed and
re-radiated by hot dust in most AGNs \cite{1989ApJ...347...29S,
1994ApJS...95....1E}. The peak of the AGN dust emission lies in the 10~$\mu$m to
20~$\mu$m range, implying temperatures of order 200~K, and reaches shortward to
$\sim$1~${\mu}$, implying T$\sim$1800~K, comparable to the maximum dust
sublimation temperature. For the simple black body case
\cite{2010ApJ...714..561L} this region spans a factor of $\sim$1000 in radius,
with most of the emitting area, and presumably most of the mass, at the larger
radii.

Indeed, direct mid-IR imaging at 12$\mu$m with VLT \cite{2008A&A...479..389H,
2009A&A...495..137H, 2008A&A...479..389H}, shows that the mid-IR emission
becomes tightly correlated with the AGN continuum only within $\sim$600
R$_{sub}$, comparable with the expected scale. More extended 12$\mu$m emission
dilutes the correlation and so presumably arises from star formation.

The dust emitting radius [$R_{dust}(T)$] can be expressed in terms of the
Eddington ratio of the accretion rate onto the black hole
(L/0.1$L_{Edd}=\lambda_{0.1}$), and the Schwarzchild radius (R$_g$):
$R_{dust}(T) = 1.5 \times 10^6 \lambda_{0.1}~L_{44}^{0.5}~T_{200K}^{-2.8} R_g$
\cite{2010ApJ...714..561L}\footnotemark.
\footnotetext{Following \cite{1987ApJ...320..537B}, for $\tau$=3,
and a typical dust grain size $a=0.05\mu m$. Here $L_{44}$ is the AGN UV
continuum luminosity in units of 10$^{44}$~erg~s$^{-1}$, and $T_{200K}$ is the
dust temperature in units of 200~K.}
Radii of about 10$^6$R$_g$ apply to the 200~K dust, while the hottest dust, at
T$\sim$1800~K, has R(hot)$\sim$3000 R$_g$.

It is instructive to compare these dust emitting radii with the size of the SMBH
sphere of influence, $R_{BHinfl}=G~M_{BH}/\sigma^2$ \cite{2005SSRv..116..523F},
so that $R_{BHinfl}=10^6\sigma_{200}~R_g$. The main AGN dust emission thus spans
the region in which the host galaxy potential takes over from the black hole
potential (Figure~1). We might expect interesting things to happen as this
boundary is crossed. A stable H/R$\sim$1 torus is unlikely to be continuous
across this region.

The most discussed alternative to explain this dust emission is a dusty wind off
the accretion disk. The wind may be driven by magneto-centrifugal forces
\cite{1982MNRAS.199..883B, 1994ApJ...434..446K,2006ApJ...648L.101E}, by external
irradiation of the disk making it unbound \cite{2005ApJ...630..167T}, or by star
formation in the gravitationally unstable part of the disk
\cite{1999A&A...344..433C}. A steady state accretion model
\cite{2004A&A...413..949V} can also sustain a large scale height torus, but only
at locally super-Eddington accretion rates.

A warped disk \cite{1989ApJ...347...29S} can have a large covering factor
without needing a large scale height, and so is an alternative to a planar
torus. This form has gained new relevance as the means by which a black hole is
fed and spun up has been investigated in more detail. Individual accretion
events with random angular momentum vectors will naturally produce a low spin
black hole \cite{2007ApJ...667..704V}. Lawrence \& Elvis
\cite{2010ApJ...714..561L} point out that a warped disk will result, and will
produce a type~1 or type~2 appearance depending on orientation, similar to that
from a donut torus, but with observable differences. Small angle warps are seen
in some AGN megamasers on parsec scales \cite{2003ApJ...590..162G}, and are
implied in the Galactic Center \cite{2008ApJ...683L..37W}.

A simple warped disk, with no rotation of the line of nodes, and random
accretion directions, predicts a 1:1 type~2:type~1 ratio
\cite{2010ApJ...714..561L}. These authors then show that this ratio is
consistent with observations, if LINERS are excluded, for AGN samples selected
by reasonably isotropic indicators (radio, [OIII], mid-IR fluxes). The exception
is samples selected in X-rays, which show a higher fraction of X-ray obscured
objects.  This suggests an extra, dust-free, absorber of X-rays, probably at
smaller radii.  These absorbers could be the broad emission line clouds noted
above (\S 3.1).

A 1:1 ratio, though, also eases several of the objections to the donut torus.
The resulting wider opening angles ease the photon and energy budget deficits,
allow wide angle biconical winds to escape, and so greater feedback, and require
a more physically reasonable smaller scale height torus. The solution is still
open.

Another prediction of the warped disk model is that the radio jets and the
bi-cones will not generally be aligned. Several anecdotal examples suggest that
this is the case \cite{2010ApJ...714..561L}, but a general survey is needed. The
canonical Hubble snapshot survey \cite{2003ApJS..148..327S} needs to be repeated
to greater depth, perhaps with AO on 8-meter class telescopes. Infrared
interferometers are now beginning to resolve this region as well as, or better
than, Hubble \cite{2009A&A...502...67Tshort, 2011A&A...527A.121K,
2009ApJ...705L..53B}, so we may be able to see warped AGN dust disks directly in
a few years. Eventually, imaging reverberation mapping will let us image the
broad line region motions too, in all three spatial and velocity components
\cite{2002ApJ...581L..67E}.

%%%%%%%%%%%%%%%%%%%%%%%%%%%%%%%%%%%%%%%%%%%%%%%%%%%%%%%%%%%%%%%%%%%%%%%%%%%
\section{Conclusions}

A complete physical model for quasars is developing, but has some way to go:
The atomic emission and absorption features all have promising explanations in
terms of winds or failed winds subject to radiation pressure
\cite{2012arXiv1201.3520E}. Although a great deal is know about the behavior of
the X-ray source, physically it is normally ascribed to a "hot corona" which is
not understood. I anticipate that accurate measurements of the temperature of
the X-ray emission by NuSTAR \cite{2010SPIE.7732E..21Hshort} will begin to
unpick this knot.

The origin of the AGN powered hot dust and of the various types of AGN is
becoming more refined, but also, perhaps, more interesting and tractable. The
view of AGN obscuration as being due to a single donut torus does not capture
the complexity now realized to exist within AGNs. While acknowledging that the
illustration from Urry \& Padovani is graphic and immediate, I would urge that
it should not be our default picture, as this form tends to become imprinted,
and is then hard to get beyond.

In the emerging, more complex, picture - as in the donut torus model - the
various types of AGNs continue to be explained by obscuration. However, at least
three separate obscuring regions must exist in quasars and AGNs: (1) an inner,
probably dust-free, region producing rapid X-ray eclipses, most likely due to
broad emission line clouds; (2) an intermediate scale dusty region that
reprocesses much of the AGN luminosity into thermal dust emission, including the
hottest dust. This region spans the boundary between the black hole- and
host-dominated gravitational potentials, and is perhaps in the form of a warped
disk; (3) outer regions on 0.1-1~kpc scales, connected to the host disk and/or
dust lanes.

By separating out the effects of these different obscurers we are starting to
learn much more about the inner structure of AGNs, and their feeding on
sub-kiloparsec scales
.

\ack The author would like to the thank his many collaborators in
this area over the years, especially Andy Lawrence and Guido Risaliti, Bozena
Czerny, and Gordon Richards, and is grateful to the scientific organizing
committee for the opportunity to present this work at AHAR 2011.

\medskip
\noindent{\bf References}
\bibliography{ELVIS}

\end{document}